\documentstyle[11pt]{article}
\newcommand\be{\begin{equation}}
\newcommand\ee{\end{equation}}
\newcommand\bea{\begin{eqnarray}}
\newcommand\eea{\end{eqnarray}}

\newcommand{\fatalpha}{{\bf \alpha \kern -0.44em \alpha}}
\newcommand{\fatsigma}{{\bf \sigma \kern -0.54em \sigma}}
\newcommand{\tpchi}{{\bf \chi \kern -0.35em \chi}}
\newcommand{\llambda}{{\bf \lambda \kern -0.45em \lambda}}

\bibliography{plain}
\title{\textbf{Randomness criteria in binary visibility graph perspective}}
\vspace{20mm} \author{ S. Ahadpour
\thanks{E-mail:ahadpour@uma.ac.ir}$\;$ , Y.
Sadra\thanks{E-mail:sadra@uma.ac.ir}
.\\
\\{\small Department of Physics, University of Mohaghegh Ardabili, Ardabil, Iran.}}
\newpage
\begin{document} \maketitle \vspace{15mm}
\begin{abstract}
By means of a binary visibility graph, we present a novel method to
study random binary sequences.  The behavior of the some topological
properties of the binary visibility graph, such as
 the degree distribution, the clustering coefficient, and the mean path
length have been investigated. Several examples are then provided to
show that the numerical simulations confirm the accuracy of the
theorems for finite random binary sequences. Finally, in this paper
we propose, for the first time, three topological properties of the
binary visibility graph as a randomness criteria.

\end{abstract}
\section{Introduction}
The relationship  between time series analysis and complex networks
have emerged \cite{N. Marwan,P. Li}. Zhang et al. introduced a
method of mapping between time series and complex networks, they
found that, the dynamics of time series are encoded into the
topology of the corresponding network \cite{Zhang J,J.-F. Sun}.
Lacasa et al. have proposed an alternative mapping between time
series and complex networks based on the visibility graph algorithm,
they are able to discriminate uncorrelated randomness
from chaos series \cite{L. Lacasa,B. Luque}.\\
 Recently, complex network theory has stimulated explosive
interests in the study of social, informational, technological and
biological systems, resulting in a deeper understanding of complex
systems  \cite{R. Albert,M.E.J. Newman,S.N.
Dorogovtsev,S.Boccaletti}. We apply visibility algorithm as a new
method for random binary sequences analysis, which converts binary
sequences into complex networks. Whereas the previous works
\cite{X.-H. Ni,Zhang J} were focused on the dynamics of a complex
system is usually recorded in the form of time series, which can be
studied through its visibility graph from a complex network
perspective. The intent of this paper is to propose a new binary
visibility graph (BVG) which stands as a subgraph of the visibility
graph.  The rest of the paper is organized as follows. In Sec.II we
introduce the BVG algorithm. In Sec.III we derive exact results for
topological properties of the BVG such as degree distribution, local
clustering coefficient, long distance visibility. we propose, for
the first time, three topological properties of the BVG as a
randomness criteria. This section is followed by an outlook section.

\section{Construction of BVG }
We start with the description of the  visibility graph. By
considering an arbitrary sampled time series $\{u_{t}:
t=1,2,...,N\}$. Each data point of the time series is encoded into a
node of the visibility graph. Two arbitrary data points $u_{i}$ and
$u_{j}$ in the time series have visibility, and consequently become
two nodes in the associated graph, if any other data point $u_{k}$
such that $i<k<j$ fulfills.
\begin{equation}
u_{k}<u_{i}+ (u_{i}-u_{j})\frac{i-k}{j-i}\;.
\end{equation}
An example of a time series containing $20$ data points and the
associated visibility graph derived from the visibility algorithm is
illustrated in (Fig.1). By definition, any visibility graph
extracted from a time series is always connected since each node
see, at least its nearest neighbors and the degree of any node
$u_{t}$ with $1<t<N$ is more than $2$. Furthermore, the constructed
graph inherits several properties of the series in its structure.
Therefore, periodic series convert into regular graphs, random
series convert into irregular random graphs and fractal series do so
into scale-free networks \cite{L. Lacasa}. It is also found that a
visibility graph is invariant under affine transformation of the
series data since the visibility criterion is invariant under
rescaling of both horizontal and vertical axes, and under horizontal
and vertical transformation \cite{D. J. Watts}.\\
The BVG is an algorithm that maps a binary sequence into a graph (as
shown in Fig.2).
 Here, we briefly describe the binary visibility algorithm in the following way:\\
  Let $\{x_{i}\}_{i=1,...,N}$ be a binary sequence of $N$ bits. The
algorithm assigns each bit of the binary sequence to a node in the
BVG the algorithm is abbreviated as BVA . Two nodes $i$ and $j$ in
the BVG are connected if one can draw a visibility line in the
binary sequence joining $x_i$ and $x_j$ that does not intersect any
intermediate bits height. $x_{i}(x_{j})$ can only be $0$ and $1$.
Therefore, $i$ and $j$ are two connected nodes if the succeeding
geometrical criterion is satisfied with the binary sequence:
\begin{equation}
x_{i}+x_{j}>x_{n} \;\;  that \;\;  x_{n}=0  \;\;for\;\; all \;\; n
\;\;such \;\;that \;\;i<n<j\;.
\end{equation}

 It is important to note that, given a binary sequence, its BVG is
a subgraph of its associated visibility graph. consequently, as in
the former case, the BVG associated with a binary sequence is always
connected and undirected, since, each node sees at least its first
neighbors (left-hand and right-hand). In what follows we will show
that the simplicity of the binary version of the algorithm allows
analytical solvability and geometrically simpler,  this new method
can attest to distinguish between random and non-random binary
sequences.

\section{ Topological properties of the BVG  }
In order to investigate some statistical characteristics of  the
binary sequences,  the following assumptions are made with respect
to random binary sequences to be tested:
\begin{description}
    \item [Uniformity:] The occurrence of zeros and ones are of
equal probabilities, i.e. if a sequence is of length $n$, the
expected number of ones (or zeros) is $n/2$.\\
    \item [Scalability:] Any subsequences should have the same
statistic characters with the sequence they randomly extracted from,
i.e.  any test applicable to a sequence can
also be applied to the subsequences.\\
    \item [Consistency:] The behavior of a generator must be consistent
across starting values (seeds).\\
\end{description}
Under above framework, The National Institute of Standards and
Technology $\left(NIST \right)$  statistical tests suite (which can
be freely down- loaded from website http://csrc.nist.gov/rng/) for
random binary sequences offers a battery of sixteen statistical
tests \cite{National}. In the following three subsections we will
present three intuitive interpretations of the topological
properties of the BVG .\\
\subsection{Degree distribution  } Let us consider a bi-infinite binary
sequence created from a binary valued random variable $X$(with $x$
as its values) such that $x\in\{0,1\}$. For  simplicity, we will
label a generic bit $x_{0}$ as the ``seed" bit here after. In order
to obtain the degree distribution $P(k)$ \cite{B. Bollobás} of the
associated graph, we are going to estimate the probability of an
arbitrary bit having  $x_{0}$ value  which can be observe, $k$ other
bits. If $k$ bits are observed by $x_{0}$, there will be encounter
with two bounding bits with  values on each side, one on the
right-hand side of $x_{0}$ and the other on its (L.H.S). So that the
$ k-2 $  visible bits  will be located in that window, i.e. they are
zeros. This implies  the minimum possible degree is $k
=2$.\\
 As these ``inner" bits should appear sorted by its position from
seed(being on the left or right side if depending in the position of
the seed), Hence we can  say that there are exactly $k-1$ different
possible configurations $\{C_{i}\}_{i=0,. . .,k-2}$, where the index
$i$ determines the number of inner bits on the right-hand side of
$x_{0}$(see Fig.3). It should be mentioned that the case where $k=4$
and $x_{0}=0$ is an exception, since the seed is always in between
two inner bits.  In this paper, for a more exacting analysis, we
study the cases $x_{0}=0$ and $x_{0}=1$, separately.
\\
We are calculated for  the first example a set of possible
configurations for a seed bit $x_{0}$ with $k=4$ result denoted in
Fig.3. As it is observe the sign of the subindex in $x_{i}$
depending   bit  is whether, it is located  at the (L.H.S)or (R.H.S)
of $x_{0}$. Therefore, the bounding's bits subindex directly
indicates the amount of bits located in that side. As an example, in
$x_{0}=1$, $C_{0}$ is the configuration where none of the $k-2=2$
inner bits are located in the (L.H.S) of $x_{0}$, and hence the left
bounding bits are labeled as $x_{-1}$ and the right bounding bits
are labeled as $x_{3}$ . For $x_{0}=0$, $C_{0}$ is the configuration
where one of the $k-2=2$ inner bits are located in the (L.H.S) of
$x_{0}$, and therefore the left bounding bits are labeled as
$x_{-2}$ and the right bounding bits are labeled as $x_{n+1}$. Note
that n hidden bits can be located in the (R.H.S) of the inner bit.
In $x_{0}=1$, $C_{1}$ is the configuration for which inner bits are
located in the (L.H.S) of $x_{0}$ and another inner bits are located
in its (R.H.S). For $x_{0}=0$, $C_{1}$is the configuration for which
$n1$ hidden bits are located in the (L.H.S) of $x_{0}$ and $n2$
hidden bits are located in its (R.H.S). Finally, in $x_{0}=1$,
$C_{2}$ is the configuration for which both inner bits are located
in the (L.H.S) of the seed. For $x_{0}=0$, $C_{2}$is the
configuration where one of the $k-2=2$ inner bits are located in the
(R.H.S) of $x_{0}$, and therefore the right bounding bits are
labeled as $x_{2}$ and the left bounding bits are labeled as
$x_{-(n+1)}$. Notice that $n$ hidden bits can be located in the
(R.H.S) of the inner bit (see Fig. 3).\\
Consequently, $C_{i}$ corresponds to the configuration for which $i$
inner bits are placed at the (R.H.S) of $x_{0}$, and $k-2-i$ inner
bits are placed at its (L.H.S). Each of these possible
configurations have an associated probability $p_{i}\equiv p(C_{i})$
 that will result in  $P(k)$ such that
\begin{equation}
\;\;\;\;\;\;\;\;\;\;\;\;\;\;\;\;\;\;\;\;\;\;\;\;\;\;\;\;\;\;\;\;\;\;\;\;\;\;\;\;\;\;P(k)=\sum_{i=0}^{k-2}p_{i}.
\end{equation}
Now, the calculation of a general relation for $P(k)$ should be done
in the following steps:\\
 In the first step, we are going to perform to calculation of Eq.(3),
 for $k=2$,i.e. the probability that the seed bits have two and only
two visible bits. These obviously will be the bounding bits that we
will label $x_{-1}$ and $x_{1}$ for (L.H.S) and (R.H.S) of the seed,
respectively. For $k\geq 2$, by taking into account the total
probability that $x_{0}$ sees is $1$. Because of any bit in the
introduced binary visibility algorithm (sec.2), sees at least its
first neighbors. Now, let us look at the particular case for Eq.(3),
taken at
 $k=2$:\\
 For $x_{0}=0$:
$$
p(x_{0}=0)= Prob(x_{1},x_{-1}=1)=\frac{1}{8}
$$

For $x_{0}=1$:
$$
p(x_{0}=1)= Prob(x_{1},x_{-1}=1)=\frac{1}{8}
$$
Then,
\begin{equation}
\;\;\;\;\;\;\;\;\;\;\;\;\;\;\;\;\;\;\;\;\;\;\;\;P(k=2)=p(x_{0}=0)+p(x_{0}=1)=\frac{1}{4}
\end{equation}
In this step, we are going to perform to calculation of  Eq.(3), for
$k=3$,i.e. for the seed which has three and only three observable
bits.
 In this process, we encounter with two different configurations : $C_{0}$, in which $x_{0}$ has
two bounding visible bits ($x_{-1}$ and $x_{2}$, respectively) and a
(R.H.S) inner bit $(x_{1}$, and the same for $C_{1}$ but with the
inner bit being placed at the (L.H.S) of the seed; so
$$P(k=3)=p(C_{0})+p(C_{1})\equiv p_{0}+p_{1}$$
Note that at this point  for $x_{0}=0$,  an arbitrary number $n$ of
hidden bits $b_{1},b_{2},\cdots,b_{n}$ can eventually be located
between the inner and the bounding bits, and this fact needs to be
taken into account in the probability calculation. The geometrical
restrictions for the $b_{j}$ hidden bits are $b_{j}=0$(
j=1,$\cdots$,n) for $C_{0}$ and $d_{j}=0$( j=1,$\cdots$,$n'$) for
$C_{1}$. Then,
$$
p_{0}(x_{0}=0)= Prob[(x_{n+1},x_{-1}=1)\cap
(\{b_{j}=0\}_{j=1,...,n})],
$$
$$
p_{1}(x_{0}=0)= Prob[(x_{-(n'+1)},x_{1}=1)\cap
(\{d_{j}=0\}_{j=1,...,n'})],
$$
  At this stage we have to consider all the hidden bits totally
  configurations ($C_{0}$ without hidden bits, $C_{0}$ with a single hidden bit,
$C_{0}$ with two hidden bits, and so on, and the same for $C_{1}$).
 With a little calculation, one obtains
\begin{equation}
p_{0}(x_{0}=0)=\frac{2}{32}[1+\sum_{n=2}^{\infty}(\prod_{j=2}^{n}p(n_{j}))]=\frac{3}{32}
\end{equation}
 where the first term in the square bracket in Eq.(5) corresponds to the contribution of a
configuration with no hidden bits and the second sums over the
contributions of $n$ hidden bits.
$$
p_{0}(x_{0}=1)= Prob[(x_{2},x_{-1}=1)\cap(x_{1}=0)]=\frac{2}{32} ,
$$

 For a similar result $p_{1}$  can be find. As a consequence of this similarity the configurations are symmetrical
 for be  $C_{0}$, $C_{1}$.  Ultimately, one gets
\begin{equation}
P(k=3)=2(p_{0}(x_{0}=0)+p_{0}(x_{0}=1))=\frac{10}{32}
\end{equation}
To continue the evaluation, we need to calculate the contributions
due to the Eq.(3), for $k=4$ , i.e. for the seed which has four and
only four observable bits. For $x_{0}=1$($x_{0}=0$), we encounter
with three different configurations: $C_{0}$, in which $x_{0}$ has
two bounding visible bits $x_{-1}$, $x_{3}$($x_{-2},x_{n+1}$)
respectively and two (R.H.S) inner bits
$x_{1},x_{2}$($x_{-1},x_{1}$) and the same for $C_{1}$ but with the
inner bits being place at the (L.H.S) of the seed; so
$$P(k=4)=p(C_{0})+p(C_{1})+p(C_{2})\equiv p_{0}+p_{1}+p_{2}$$
Note at this point that for $x_{0}=0$,  an arbitrary number $n$ of
hidden bits $b_{1},b_{2},\cdots,b_{n}$ can eventually be located
between the inner and the bounding bits, and this fact needs to be
taken into account in the probability calculation. The geometrical
restrictions for the $b_{j}$($b_{i}$) hidden bits are
$b_{j}=0$(j=1,$\cdots$,n2)[$b_{i}=0$( i=-1,$\cdots$,-n1)] for
$C_{0}$ and the same for $C_{1}$,$C_{2}$. Then,

$$
p_{0}(x_{0}=0)= Prob[(x_{n2+1},x_{-(n1+1)}=1)\cap
(\{b_{j}=0\}_{j=1,...,n2})\cap(\{b_{i}=0\}_{i=-1,...,-n1})],
$$
Now, we need to consider every possible hidden bits configuration
($C_{0}$ without hidden bits, $C_{0}$ with a single hidden bit,
$C_{0}$ with two hidden bits, and so on, and the same for
$C_{1}$,$C_{2}$). With a little calculation, one obtains
\begin{equation}
p_{0}(x_{0}=0)=\frac{1}{32}[1+2\sum_{n=2}^{\infty}(\prod_{j=2}^{n}p(n_{j}))]=\frac{2}{32}
\end{equation}
where the first term in the square bracket in Eq.(7) corresponds to
the contribution of a configuration with no hidden bits and the
second sums over the contributions of $n1$ and $n2$ hidden bits.
$$
p_{0}(x_{0}=1)=
Prob[(x_{3},x_{-1}=1)\cap(x_{1}=0)\cap(x_{2}=0)]=\frac{1}{32} ,
$$
We obtain similar results for   $p_{1}$($p_{2}$) and consequently
the configuration provided by $C_{1}$($C_{2}$) is symmetrical to the
one provided by $C_{0}$.  Ultimately, one gets
\begin{equation}
P(k=4)=3(p_{0}(x_{0}=0)+p_{0}(x_{0}=1))=\frac{9}{32}
\end{equation}
Let us proceed by tackling the case $P(k=5)$, that is, the
probability that the seed has five and only five visible bits. Four
different configurations arise: $C_{0}$, in which $x_{0}$ has two
bounding visible bits $x_{-1}$, $x_{4}$ respectively and three
right-hand side inner bits $x_{1},x_{2},x_{3}$ and the same for
$C_{1},C_{2},C_{3}$ but with the inner bits being place at the
left-hand side of the seed; so
$$P(k=5)=p(C_{0})+p(C_{1})+p(C_{2})+p(C_{3})\equiv p_{0}+p_{1}+p_{2}+p_{3},$$
Then,
$$
p_{0}(x_{0}=1)=
Prob[(x_{4},x_{-1}=1)\cap(x_{1}=0)\cap(x_{2}=0)\cap(x_{3}=0)]=\frac{1}{64},
$$
We can find an identical result for $p_{1}$($p_{2}$,$p_{3}$) and
consequently the configuration provided by $C_{1}$($C_{2}$,$C_{3}$)
is similar to the one provided by $C_{0}$.  Ultimately, one gets
\begin{equation}
P(k=5)=4p_{0}(x_{0}=1)=\frac{4}{64}
\end{equation}
The results of the present calculations are summarized:

$$
 P(x_{0}=0)= \left\{
 \begin{array}{l} \frac{4}{32}\;\;\;\;\;\;\;\;k=2\\
\frac{6}{32}\;\;\;\;\;\;\;\;k=3 ,4\\
 0\;\;\;\;\;\;\;\;\;\;k\geq5\\
     \end{array}\right.
 $$

$$P(x_{0}=1)=\frac{(k-1)}{2^{k+1}}\;\;\;\;\;\;\;\;k\geq2$$

Therefore, we can argue that, for $k\geq5$:
\begin{equation}
\;\;\;\;\;\;\;\;\;\;\;\;\;\;\;\;\;\;\;\;\;\;\;\;\;\;\;P(k)=\frac{(k-1)}{2^{k+1}}
\end{equation}
But, in general,
\begin{equation}
\;\;\;\;\;\;\;\;\;\;\;\;\;\;\;\;\;\;\;\;\;\;\;\;P(k)=P(x_{0}=0)+P(x_{0}=1)
\end{equation}
 We can achieve that, the degree distribution $P(k)$ of the associated BVG
 has the semiexponential form.\\
The values of $\chi^{2}$ goodness-of-fit test between the
theoretical prediction degree distribution Eq. (11) and numerical
results demonstrated the measure of uniformity. In order to confirm
further the accuracy of our analytical results for the case of
finite binary sequences, we have performed several numerical
simulations. We have generated random binary sequences of $10^{6}$
bits and  their associated BVG. In Fig. 4 we have plotted the degree
distribution of the resulting graphs (triangles correspond to a
sequence extracted from a CCCBG tent map \cite{Narendra K }, while
circles correspond to one extracted from a CCCBG logistic map
\cite{Narendra K ,Ali Kanso}, respectively). The line is the best
fit of the theoretical , showing a perfect agreement with the
numerics.

\subsection{Local clustering coefficient distribution }
By means of geometrical arguments, we can obtain the local
clustering coefficient $C$ \cite{M.E.J. Newman,R. Albert,S.
Dorogovtsev,S. Boccaletti,B. Bollobás} of a BVG associated with a
binary sequence. For a reference node $i$, $C$ means the rate of
nodes connected to $i$ that are connected between each other, where
$C$ represents the clustering. In other words, we have to work out
from a reference node $i$ how many nodes from those visible to $i$
have mutual visibility (triangles), normalized with the set of
possible triangles $(^k_{2}) $. In a first step, if a generic node
$i$ has degree $k=2$, these nodes are straightforwardly two bounding
bits, hence having mutual visibility. Hence , in this condition
there exists one triangle and $C(k=2)= 1$. Now if a generic node $i$
has degree $k=3$($k=4,5$), one (two,three) of its neighbors will be
an inner bit(two, three bits), which will only have visibility of
one of the bounding bits (by construction). We achieve that in this
condition we can only form three (five,five) triangles out of
three(six,ten) possible ones, thereby:
\begin{equation}
C(k)= \left\{
 \begin{array}{l} 1 \;\;\;\;\;\;\;\;\;\;\;\;k=2,3\\
\frac{5}{6}\;\;\;\;\;\;\;\;\;\;\;\;k=4\\
 \frac{2k-5}{(^k_{2})}\;\;\;\;\;\;k\geq5\\
     \end{array}\right.
\end{equation}
This relation between $k$ and $C$ for $k\geq5$ allows us to deduce
the local clustering coefficient distribution $P(C)$ as follows:
$$
P(k)=\frac{k-1}{2^{k+1}}=P\left(\frac{f(C)-C+4}{C2^{\frac{5C+4+f(C)}{2C}}}\right),
$$
Where $f(C)=(C^{2}-32C+16)^{\frac{1}{2}}.$ In general,
\begin{equation}
P(C)= \left\{
 \begin{array}{l} \frac{f(C)-C+4}{C2^{\frac{5C+4+f(C)}{2C}}} \;\;\;\;\;\;\;\;\;0< C \leq\frac{5}{10}\\
\frac{9}{32}\;\;\;\;\;\;\;\;\;\;\;\;\;\;\;\;\;\;\;\;\;\;\;\;\;\;\;\;\;C=\frac{5}{6}\\
 \frac{18}{32}\;\;\;\;\;\;\;\;\;\;\;\;\;\;\;\;\;\;\;\;\;\;\;\;\;\;\;\;\;C=1\\
     \end{array}\right.
\end{equation}

To confirm the validity of this latter relation within finite binary
sequences, in Fig. 5 we illustrate the clustering distribution of a
BVG associated with a random binary sequence of $10^{6}$ bits
(circles) obtained numerically. The line is the best fit of the
theoretical and triangles corresponds to the theoretical prediction
$(C=1,\frac{5}{6})$, in excellent agreement with the numerics.\\
 The values of $\chi^{2}$ goodness-of-fit test between the theoretical
prediction
 clustering distribution Eq. (13) and numerical results
demonstrated the measure of consistency.
\subsection{Long distance visibility, mean degree, mean path length  }
The mean path length scaling   \cite{B. Bollobás}, can be derived as
below, let us first estimate the probability $P(n)$ that two bits
separated by $n$ intermediate bits be two connected nodes in the
graph. By taking into account a binary sequence to construct
associated BVG. An arbitrary  $x_{0}=1$ from the mentioned sequence
can be  ``observe" $x_{n}=1$ (and therefore would be connected to
node $x_{n}$ in the graph) if and only if $x_{i}=0$ for all $x_{i}$
$(i=1,2, . . . ,n-1)$. Then $P(n)$ may  be estimated as
\begin{equation}
P(n)=
prob[(x_{0},x_{n}=1)\cap(\{x_{i}=0\}_{i=1,...,n-1})]=\frac{1}{2^{n}}
\end{equation}
Now, we can derive the mean degree $<k>$ of the binary visibility
graph as follows:
\begin{equation}
<k>=\sum kP(k)=3.5,
\end{equation}
which we can be obtained from $P(n)$ as
\begin{equation}
<k>=3.5\sum _{n=1}^{\infty}P(n)=3.5.
\end{equation}
At this point, in the Fig. 6 to illustrate  the adjacency matrix
\cite{B. Bollobás} of the BVG associated with a random binary
sequence of $500$ bits ( if nodes $i$ and $j$ are connected, then
the entry  $i$ , $j$ are filled in black and  otherwise they are
filled, blank  ). Since every bit $x_{i}$ has visibility of its
first neighbors $x_{i-1}$ ,$x_{i+1}$, every node i will be connected
by construction to nodes $i-1$ and $i+1$: the graph is thus
connected. The  Fig. 6 indicates that the graph is very to exact
 homogeneous structure,i.e. the adjacency matrix is exactly
filled around the main diagonal. Moreover, the matrix evidences a
superposed compact structure, noticeably the visibility probability
$P(n)=\frac{1}{2^{n}}$ that introduces some shortcuts in the BVG,
much in the vein of the small-world model \cite{D. J. Watts}. Here,
the $P(n)$ denotes,  the shortcuts probability.From the Statistical
point of view, we can interpret the graph's structure as nearly
homogeneous, where by increasing the size of graph's, the size of
the local neighborhood do not change. Hence, we can approximate its
mean path length $L(N)$ as
\begin{equation}
L(N)\approx
\sum_{n=1}^{N-1}nP(n)=\sum_{n=1}^{N-1}\frac{n}{2^{n}}=2(1-\frac{N+1}{2^{N}})
\end{equation}
It is observe that, the logarithmic scaling emerged , denoting that
the BVG associated with a generic random sequence is small world
\cite{D. J. Watts},  which may be observed in the Fig. 5. The
numerical results of $L(N)$ (circles) of a BVG associated with
several random binary sequences of increasing size $N=2^{7},2^{8} ,
. . .,2^{19}$ in the Fig. 7, have been plotted . The line is the
best fit of the theoretical. The values of $\chi^{2}$
goodness-of-fit test between the theoretical prediction mean path
length Eq. (17) and numerical results demonstrated the measure of
scalability.\\

\section{Conclusion and outlook}
 In this article, we have investigated the binary visibility graph, constructed from
the random binary sequences. The present study illustrates the
uselessness of the previous works in the analysis of random binary
sequences \cite{A.-L. Barabási,Chuang Liua,B. Luque}. We have also
evaluated exact results on several topological properties of the BVG
associated with generic uncorrelated random binary sequences, and
numerical simulations confirmed its reliability for finite
sequences, and the results show the three topological properties of
the binary
visibility graph as a excellent randomness criteria.\\
 Furthermore, we do hope that our obtained results through this paper will pave the
way for further studies on nonlinear dynamical systems.

\section{Acknowledgments}
The authors would like to express their heartfelt gratitude to Mr.
D. Manzoori, Mr. S. Behnia for the nice editing of their paper.
 %for the nice editing of their paper - this has
%certainly improved its readability.

\newpage


\begin{thebibliography}{99}
%Intro
\bibitem{N. Marwan} N. Marwan, J. F. Donges, Y. Zou, R. V. Donner, J. Kurths, Complex
network approach for recurrence analysis of time series, Phys.Lett.A
373 (2009) 4246-4254.
\bibitem{P. Li} P. Li,  B. H. Wang, An approach to Hang Seng Index in Hong Kong stock
market based on network topological statistics, Chinese Science
Bulletin 51 (2006) 624- 629.
\bibitem{Zhang J}  J. Zhang , M. Small,   Complex network from pseudoperiodic time series: Topology
versus dynamics. Phys Rev Lett 96, (2006), 238701- 238704.
\bibitem{J.-F. Sun} J. Zhang, J. F. Sun, X. D. Luo, K. Zhang, T. Nakamura, M. Small,
 Characterizing pseudoperiodic time series through the complex network
approach, Physica D 237 (2008) 2856– 2865.
\bibitem{L. Lacasa} L. Lacasa, B. Luque, F. Ballesteros, J. Luque, J. C. Nuno,
 From time series to complex networks: The visibility graph, PNAS,
105, (2008) 4972-4975.
\bibitem{B. Luque} B.  Luque,  L.  Lacasa, F. Ballesteros,  J. Luque,
  Horizontal visibility graphs: Exact results for random time series,
 physical review E 80,  (2009) 046103- 11.
\bibitem{R. Albert} R. Albert, A. L. Barabasi,   Statistical mechanics of complex networks, Rev Mod Phys
74, (2002), 47–97.
\bibitem{M.E.J. Newman} M. E. J. Newman,  The structure and function of complex networks, SIAM Rev
45, (2003) 167–256.
\bibitem{S.N. Dorogovtsev} S.N. Dorogovtsev, J. F. F. Mendes, Evolution of Networks: From Biological Nets to the
Internet and the WWW, Oxford University Press, Oxford, 2003.
\bibitem{S.Boccaletti} S. Boccaletti, V. Latora, Y. Moreno, M. Chavez, D. U. Hwang, Phys. Rep.
424 (2006) 175- 308.
\bibitem{X.-H. Ni} X. H. Ni, Z. Q. Jiang, W. X. Zhou,
 Degree distributions of the visibility graphs mapped from fractional Brownian motions and multifractal
random walks, Phys. Lett. A, 373 (2009) 3822 -3826.
\bibitem{D. J. Watts}  D. J. Watts and S. H. Strogatz, Collective dynamics of 'small-world' networks,
 Nature,  393, (1998) 440 -442.
 \bibitem{National} National Institute of Standards and Technology, A
statistical testsuit for random and pseudorandomnumber generators
for cryptographic applications, NIST special publication, (2001)
800-22.
\bibitem{B. Bollobás}  B. Bollobás, Modern Graph Theory, Springer-Verlag, New
York, (1998).
\bibitem{Narendra K }  P.  K.  Narendra ,  P.  Vinod , K. K.
Sud,  A Random Bit Generator Using Chaotic Maps, International
Journal of Network Security, 10, (2010) 32-38 .
\bibitem{Ali Kanso}  A. Kanso, N. Smaoui, Logistic chaotic maps for binary numbers
generations,Chaos, Solitons and Fractals, 40 (2009) 2557 –2568.
\bibitem{S. Dorogovtsev}  S. Dorogovtsev,  J. F. F. Mendes, Evolution of networks,
 Adv. Phys. 51, (2002) 1079-1187 .
\bibitem{S. Boccaletti}  S. Boccaletti, V. Latora, Y. Moreno, M. Chávez,
  D. U. Hwang, Complex networks: Structure and dynamics, Phys. Rep. 424,(2006) 175-308.
\bibitem{A.-L. Barabási}  A. L. Barabási, R. Albert, Emergence of scaling in random networks,
Science 286 (1999) 509 -512.
\bibitem{Chuang Liua} L. Chuang , Z. Wei-Xing , Y. Wei-Kang ,
Statistical properties of visibility graph of energy dissipation
rates in three-dimensional fully developed
turbulence,http://arxiv.org/abs/0905.1831v2.
%\bibitem{17}




%\bibitem{M. E. J. Newman}M. E. J. Newman, SIAM Rev. 45, 167 (2003).




 \end{thebibliography}
\end{document}